# An IoT-Enabled Smart Home Automation System for Energy Efficiency with Web-Based Control

Amaan Ahmed[1], Mohammed Mahir Rahman[1][0009-0007-4167-1475], Shahzad Memon[1][0000-0003-3354-5798] and Tauseef Ahmed[1][0000-0002-1850-3496]

[1] University of East London, London E16 2RD, UK
T.Ahmed4@uel.ac.uk

**Abstract.** This paper illustrates the design and implementation of a smart home automation system for the conservation of energy and user control with the help of the environmental sensors and Raspberry Pi 5. It monitors real-time conditions like motion, temperature, humidity, light and smoke to automatically control the device's behavior and save energy. A prototype single two-room was developed which uses GPIO/I2C interfaces to integrate sensors and actuators. The fan speed and LED brightness was dynamically controlled using PWM. Manual control and real-time monitoring are made possible through a web dashboard that was developed using Flask and graphical displays, and CSV logs of the energy are taken every 30 seconds. It was designed in an iterative model of sprints and the energy savings during testing was more than 46% over an al-ways-on model. The results prove that with the help of these low-cost, modular de-vices it is possible to improve sustainability and usability in the home as part of the IoT.

## 1 Introduction

With the development of the Internet of Things (IoT), the way residential homeowners interact with their homes is changing. environments are managed, providing new opportunities for automation, energy efficiency, and user convenience. In the realm of smart home, IoT is used to refer to the integration. Network of interconnected devices, sensors and communication technologies that allow for Control equipment to monitor and control lighting, heating, ventilation, etc. of domestic systems in real time, and appliances [1]. This paradigm shift has been made because of the growing global increase in the demand for high-quality communication and its skills. Climate change awareness, cost of energy and the need for sustainable living solutions. consumption, their environmental impact is also substantial. Households make up a large share of total energy consumption worldwide, and hence have a considerable impact on the environment. The home automation, which is able to develop into an intelligent, responsive and energy-efficient house, is the consuming sector. The consuming sector is the home automation which can evolve into an intelligent, responsive, and energy-efficient home. Research and innovation on systems has turned into an important field [2]. Despite the While smart home technologies have been promised, many of the solutions that are available today are expensive, closed source. Is not competent or adaptable. There are still a significant number of households who are dependent on these. on manual controls, resulting in unnecessary energy waste due to appliances and lighting used, Unoccupied areas have the heaters turned on. This study has repeatedly demonstrated the benefits of integrating. With automation logic, an energy savings of 20-50% can be achieved through environmental sensing [3, 4, 5]. But there are challenges regarding its affordability, scalability, and user accessibility. These are the gaps this studies will fill by the design and implementation of a Smart Home Automation System Based on low cost and IoT which integrates environmental sensing, automatic control and easy-to-use web-based interface for real-time monitoring. Manual override and monitoring. This work is motivated by an observation that: that a lot of household energy waste is a matter of simple human negligence, Leaving unoccupied rooms with the lights or fans on. Automating the operation of the use of appliances that respond to the environment will be a practical way to help lower such. Save waste, reduce utility expenses and help fulfil sustainability objectives. Prior studies have shown that, with threshold triggered automation, devices are activated when the threshold is reached and then turned off when it is not. Under certain environmental conditions, can drastically decrease unnecessary energy usage [3]. The more advanced techniques, like predictive control, and neural networks were also employed. Several other approaches, such as the use of and fuzzy logic, have been proven to achieve a balance between user comfort and energy savings [6] While the availability of cheap embedded platforms like the Raspberry Pi has grown. Pi has helped researchers and practitioners to test low-cost smart house prototypes. systems. They can be integrated with various types of sensors such as There are many different types of sensors available, such as motion detectors, temperature sensors, humidity sensors, and light sensors, that can be incorporated. It used to start context-aware automation [7], [8]. The use of open-source software with a lightweight framework like Flask, it's possible to create lightweight, web-based dashboards. which give people instant access into their house and control [9] and [10]. To overcome these challenges, this study created a modular, low-cost and locally controlled smart home automation system. Using a Raspberry Pi 5 as the central, the system was able to connect several environmental sensors such as PIR motion sensor to controller. detectors, DHT22 temperature and humidity sensors, BH1750 light sensors and smoke detectors are employed. Detectors to make fans and LED lights automatic. Pulse Width Modulation To dynamically vary the speed of the fans and brightness of the LEDs, (PWM) was used. They found that energy use was directly proportional to environmental factors. A Flask-based web dashboard contained real-time monitoring, manual override and energy usage logging increases transparency and user control.

This paper makes the following contributions:

1. A summary of the existing techniques and technologies in residential and commercial IoT automation systems

2. An assessment of the effects of these systems on energy efficiency and sustainability.
3. Development of the prototype smart home system with Raspberry Data from pi 5 and corresponding sensors.
4. Creation of user-friendly Web interface for real-time monitoring. and device control
5. Assessment of adaptability and scalability of the system to reality deployment
6. Security concerns of IoT devices in smart home
7. Evaluation of the environmental benefits of energy saving through IoT automation.

## 2 Related Work

Smart home systems in the IoT world consist of interconnected sensors, devices and communication protocols, providing automation, comfort, safety and energy-efficiency within the home environment, enhanced by layered IoT architectures (edge, fog, cloud) which make things respond faster and decentralise decision-making such as, for example, local actuation for lighting and HVAC [1]. For building automation and real-time control and status feedback, publish-subscribe messaging using MQTT enables low bandwidth communications; ZigBee, Wi-Fi, MQTT, CoAP and 6LoWPAN trade-offs are highlighted in comparative anal-yses for low power, multi-sensor deployments [13]. Four-layer designs that combine perception, processing, edge control, and cloud storage, along with lighter protocols, directly impact energy efficiency, responsiveness and scalability, and represent a general shift towards embedded security, distributed computing and AI in third-generation IoT systems [12]. In fact, extensive research on energy efficiency has shown that threshold triggered control (TTC) with occupancy and environmental sensing can save between 20% and 50% of energy, with simulation results indicating that thresholds for HVAC and lighting can be activated only under certain conditions resulting in up to 40% energy savings [2] [3]. The incorporation of real time pricing with Smart Energy Management Systems has been shown to achieve significant reductions in consumption and cost through dynamic appliance scheduling of deferrable loads, and the retrofits in multi-storey buildings have reported 30-50% savings, with payback periods of around 6.2 years, highlighting the economic as well as sustainability benefits. In addition to rule-based control in basic HVAC systems [3] [6] user comfort and energy minimization is further facilitated by predictive and fuzzy logic controllers that adjust HVAC settings by making use of past and current sensor data. A series of Raspberry Pi, Arduino and ESP/NodeMCU embedded hardware integrations shows low cost, modular prototypes for appliance control and environmental monitoring with low latency actuation without external cloud systems [7], [8]. Raspberry Pi's capability to run multiple tasks and support a web interface is strong, while the efficient processing capabilities of NodeMCU make it an ideal choice for interfacing with the physical world and its own native Wi-Fi module, and Arduino's deterministic I/O meets demands for precise sensing makes it a good option in the processing layer for hybrid designs [12],[17]. Though practical deployments focus on sensor calibration, firmware compatibly-ship, local preprocessing to reduce network load, and multi-modal feedback (visual and sound) to improve responsiveness and user awareness in the home, there are other key issues as well. Usability is key, with Web-based dashboards, and Flask allowing for lightweight, RESTful integration of Python-based control and real-time visualization, and multi-level dashboards from households up to utility stakeholders through MQTT and analytics [9], [11]. For medium scale systems, Flask is suitable, but for bigger deployments, scalability issues have motivated the use of Flask in combination with MQTT/Node.js, and real-time graphing with Chart.js or Flot.js enables live monitoring of the system and tracking of its energy use for behavioural feedback and transparency [10], [13]. To summarise, previous studies demonstrate good advances in sensing-driven automation, energy analytics, and web control, but there are still challenges regarding the adoption of a unified, cost-effective, and retrofit-friendly system architecture that combines real-time energy monitoring, adaptive automation, user-friendly dashboards, and usable security in typical homes [11], [16]. This study aims to provide a modular, affordable, locally controlled and extensible multi-sensory input (motion, temp, humidity, light, smoke) system with PWM control coupled to Flask web dashboards and CSV energy logging to support energy-saving demonstration and best practices, as outlined in [10], [12].

# 3 Methodology

The development step used a design–build–evaluate approach similar to the engineering research process. The process started with the identification of requirements, based on the literature review, then iterative prototyping and quantitative testing. A positivists approach was used, with sensor readings, PWM duty cycles, and logged energy consumption being used to validate the system's performance.

The development was done incrementally, with the development model being agile, breaking down each sprint. The following subsystems were a primary focus of these sprints: sensor integration, actuator control, backend development, dashboard design and energy logging. Testing, refining and improving the system then moving on after each sprint. This was a continuous cycle which provided flexibility, quick de-bugging and continuous improvement.

System Implementation

1. Hardware Design

The system was built on a Raspberry Pi 5 as the central controller, with two breadboards representing two rooms. Each room contained:

- Sensors: PIR motion detector, DHT22 temperature/humidity sensor, BH1750 light sensor, MQ-series smoke sensor (Room 2).
- Actuators: PWM-controlled fan, dimmable LED, buzzer.

- Supporting Components: RTC module, resistors, jumper wires, regulated power rails.

The Raspberry Pi interfaced with sensors and actuators via GPIO and I²C. The BH1750 and RTC shared the I²C bus, while PWM-enabled GPIO pins controlled LEDs and fans. Pin mapping was carefully structured: GPIO13 and GPIO19 for fans, GPIO27 and GPIO35 for LEDs, GPIO21 for smoke detection, and GPIO22/23 for buzzers. Common grounding and regulated voltage ensured stable operation.

2. Software Stack

The software was implemented in Python 3, with Flask providing the backend web server. Key modules included:

- RPi.GPIO and gpiozero for GPIO and PWM control.
- Adafruit_DHT for DHT22 readings.
- smbus2 and adafruit_bh1750 for I²C communication.
- threading for concurrent execution of automation, logging, and dashboard updates.
- csv and json for structured data logging and configuration persistence

The Flask dashboard was built with HTML, CSS, Bootstrap, and JavaScript. AJAX calls fetched live sensor data, while Chart.js rendered real-time graphs of temperature, humidity, light, and energy usage. The dashboard supported AUTO, MANUAL, ON, and OFF modes, with PWM sliders available in manual mode.

3. Automation Logic

Automation was implemented using threshold-based rules combined with PWM scaling:

- LEDs: Brightness scaled linearly with lux values (0% duty cycle above 2000 lux, 100% below 100 lux).
- Fans: Speed scaled with temperature and humidity (40% duty cycle above 24 °C, 70% above 27 °C, 100% above 30 °C with humidity >70%).
- Motion: PIR detection acted as a gating condition, ensuring devices only operated when rooms were occupied.
- Smoke: MQ sensor triggered both room buzzers simultaneously.

These rules were executed in a continuous automation loop, applying PWM values directly to GPIO outputs. The automation logic was implemented as follows:

```
if lux > 2000:
led_pwm.ChangeDutyCycle(0)
elif lux < 100:
led_pwm.ChangeDutyCycle(100)
else:
duty = int((2000 - lux) / 20)
led_pwm.ChangeDutyCycle(duty)
```

4. Data Logging and Monitoring

The estimated energy consumption was based on PWM duty cycles and rated wattages (9W LED, 50W fan). The power consumption was measured as it happened, converted to kWh and then the power costed at £0.34/kWh. Every 30 seconds, logs were added to energy_log.csv containing the values of the sensors, the PWM states, and cost. The dashboard provided live updates and CSV reports for further analysis.

5. Validation and Testing

Differences in the environment were always responded to within 1-2 seconds by the system and the data recorded indicated that the environmental conditions were the same as were the behaviour of the actuators.

- Unit Testing: All the sensors and actuators are checked separately in the Unit testing.
- Integration Testing: End to End pipeline tested from the sensor input to actuator response and dashboard status
- Functional Testing: Manual override was successful and automated was overruped by functional testing.
- Performance Testing: Response times, logging accuracy and energy savings proven in different scenarios.

B. Ethical Considerations

This study did not include any human subjects and did not gather or analyse any personal or sensitive information. In all experiments, physical sensors alone were used in controlled environments. Audio and video recordings were not audio or video recorded throughout the testing process. The system was fully local, without relying on third-party cloud services. This design strategy reduced the potential privacy and security concerns and ensured that all data was completely within the control of the research team.

## 4 Experimental Setup

The experimental evaluation was performed on Raspberry Pi 5 with Raspberry Pi OS is based on Python 3.11. Two breadboards were used to model two rooms, each the kits include a DHT22 temperature/humidity sensor, a BH1750 light sensor, a PIR A PWM controlled fan, a dimmable LED, A motion detector, and a buzzer. Room 2 additionally the circuit features a smoke sensor using the MQ-series type, plus an RTC module for providing time. synchronization. All components were supplied by regulated 5 V supply, GPIO Communication via and I²C interfaces. The system was accessed locally via a Flask and was launched on the Raspberry Pi's dashboard. The automation thresholds were set to represent realistic household scenario. LEDs were dimmed above 2000 lux and were full brightness below 100 lux, with PWM scaling between. The duty cycle of the fans was 70% at temperatures above 24 °C and 40% below 24 °C. above 27 °C, and 100% above 30 °C when humidity exceeded 70%. PIR motion detection, the smoke sensor was used as a gating condition for both lights and fans and the smoke sensor used as a gating condition worked when the smoke was present. Activates two room buzzers at once.

The readings from the sensors were changed every second, the log of the and energy usage was recorded at 30-second intervals, and energy usage was estimated from a log of the same data. PWM duty cycles and rated wattages (9 W LEDs, 50 W fans) and costed at £0.34/kWh. The tests were carried out under controlled conditions to simulate common domestic situations. These included occupancy and non-occupancy cycles, variations in ambient When it comes to smoke alarms, these are the factors that impact the performance of smoke alarms: light, temperature and humidity changes, smoke detection events, and manual override. through the dashboard. The scenarios were repeated several times to ensure consistency.

To measure the energy saving two modes were compared: an always-on configuration where both LEDs and fans were continuously functioning, and the smart system operating. under automation logic. The energy consumption, cost, and duty cycle distributions were Measured for a 12-hour period of a test.

Following metrics were gathered:

- Response time between Average Duty cycles,
- Total Energy used,
- sensor triggers and actuator response and
- Correlation of dash and CSV logs.

## 5 Results and Evaluation

The prototype system successfully integrated all sensors, actuators, and the Flask dashboard. Smoke detection activated both room buzzers simultaneously (shown in *Fig. 1*). Motion sensors correctly gated LED and fan activation, light sensors adjusted LED brightness via PWM scaling, and temperature/humidity thresholds triggered fan speed adjustments (shown in *Fig. 2*). The dashboard also displayed real-time sensor values, supported AUTO/MANUAL/ON/OFF modes, and allowed CSV export of energy logs (shown in *Fig. 2*).

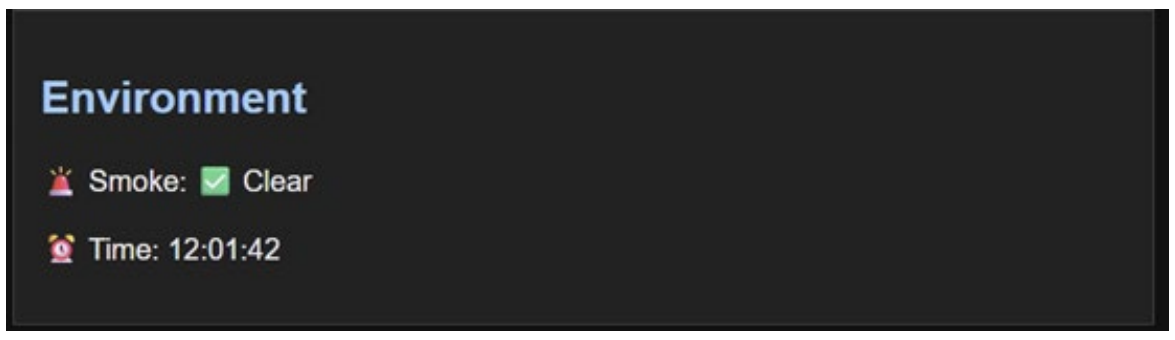

**Fig. 1.** Smoke Detector Status

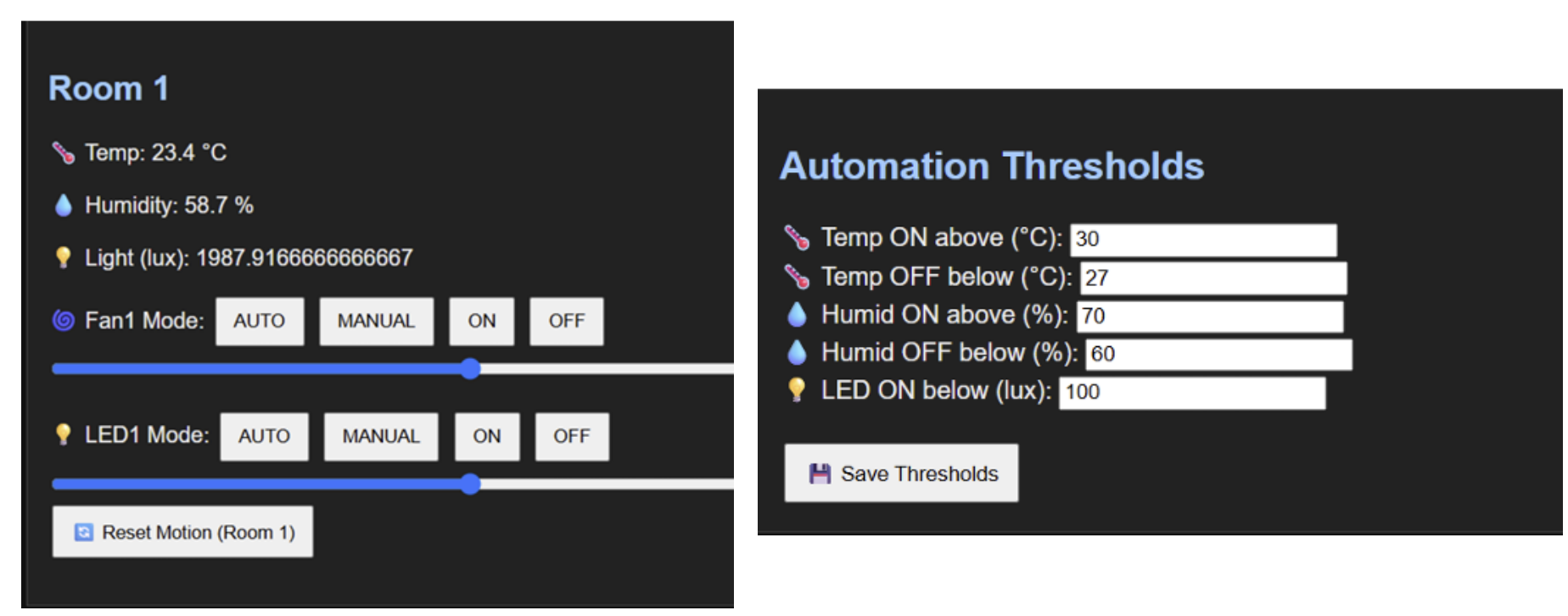

**Fig. 2.** Flask Dashboard Snippet for Room 1 (left) and User Based Thresholds (right)

### 5.1 Energy Consumption Analysis

A 12-hour test compared the always-on baseline with the smart automated system. Results demonstrated significant energy savings as shown in *Table 1*.

**Table 1.** Energy Consumption and Cost Comparison (12-Hour Test)

| Configuration | Always On | Smart System |
|---|---|---|
| **LED Energy (kWh)** | 0.216 | 0.097 |
| **Fan Energy (kWh)** | 1.200 | 0.660 |
| **Total Energy (kWh)** | 1.416 | 0.757 |
| **Cost (£)** | 0.481 | 0.257 |
| **Savings (%)** | - | **46.5%** |

The smart system reduced energy consumption by approximately 46.5%, validating the effectiveness of motion gating and PWM scaling.

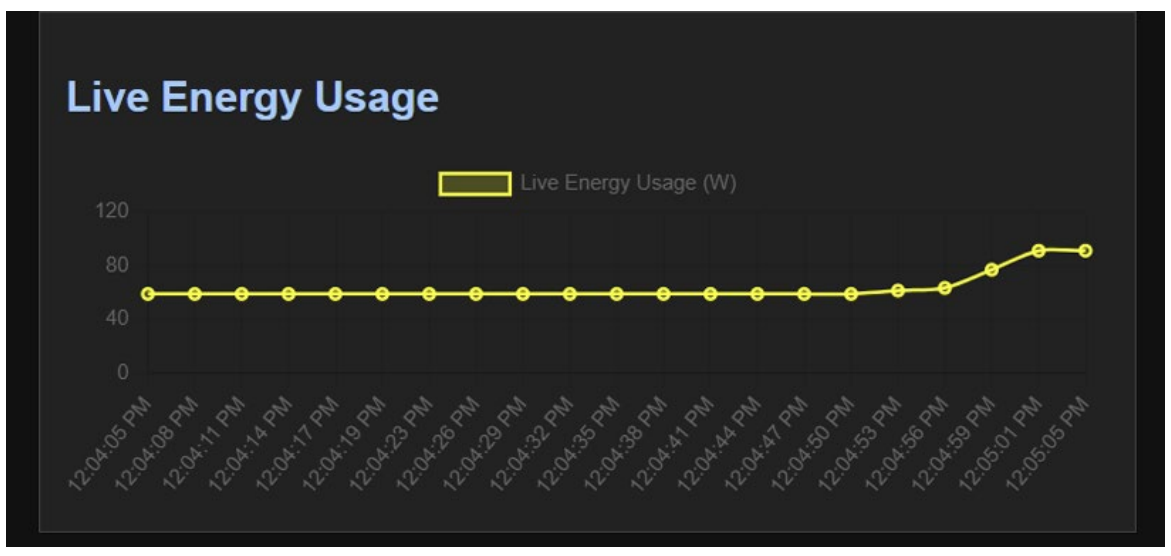

**Fig. 4.** Energy Usage Line Chart

The system logged sensor values, PWM duty cycles, and energy usage every 30 seconds. CSV logs provided timestamped records, while Chart.js graphs displayed live temperature, humidity, lux, and energy usage trends. Data consistency was confirmed between dashboard readings and CSV logs.

### 5.2 Performance Metrics

The system performed well in all of the test scenarios. Response time from the sensor triggers to the responses of the actuators was 1-2 seconds. The refresh rate of the dashboard was 1 second and the logging period was set to 30 seconds. A continuous 12-hour run confirmed system stability (shown in Table 2).

**Table 2.** Energy Consumption and Cost Comparison (12-Hour Test)

| Metric | Result |
|---|---|
| Response Time | 1-2 seconds |
| Logging Interval | 30 seconds (consistent) |
| Dashboard Refresh | 1 second |
| System Uptime | 12 hours (stable) |

### 5.3 Evaluation and Discussion

The results have shown that the system met its goals of energy efficiency, functionality and modularity. The dashboard was intuitive to control and transparent, and the energy savings were comparable to the ~46% reported in previous research. The modular design supports and make scalability easier to more rooms or appliances with minimal modification.

Limitations include reliance on breadboard wiring, which is impractical for permanent use, and the slight lag on the dashboard when CPU usage is high. The security features were limited and only included simple login/logout. The limitations of these are covered in the conclusion and future work section.

### 5.4 Security Vulnerabilities and Exploitation Risks

A thorough examination of the system identifies several vulnerabilities that adversaries could exploit to compromise the system. The greatest threats are the web interface, which is currently based on simple login/logout without hardened session management. This exposes the system to brute force attacks, credential theft and session hijacking, which could lead to unauthorized access to controls on the device. Likewise, there may be APIs that accept unauthenticated requests and/or APIs with weak validation that allow an attacker to feed data into the actuator and cause them to behave differently (e.g., forcing a constantly high duty cycle) or to cause some disruption to the normal functioning.

The absence of transport layer encryption further increases exposure, as all HTTP traffic including credentials and session tokens are sent over the web in plain text and can be read or altered by an attacker (insider threat) on the local network. Furthermore, CSV logs and configuration files are not integrity protected and can be modified to hide malicious activity or to manipulate energy usage data.

Overall, the current design is working well but it is susceptible to exploitation via weak authentication, insecure communications, unvalidated inputs, and inadequate logging protection. These risks could be used to gain unauthorized control and to manipulate device states and/or compromise data integrity and availability. To mitigate these vulnerabilities, it is necessary to create more robust authentication and session management, encrypted communication, robust input validation, integrity controls on logging, and least privilege execution of the service.

## 6 Conclusion

The tasks of the studies were accomplished successfully by designing and implementing a smart home automation system, which is modular and based on the principles of IoT. A Raspberry Pi 5 controller connected to several environmental sensors was used to control the fans and LED lighting (via PWM scaling), and a Flask dashboard was used to provide real-time monitoring, manual override, and energy logging. Experimental testing showed that the system achieves a reduction of about 46.5% in energy consumption compared to a reference case that has the system always on, proving the efficiency of the motion gating and adaptive control. The studies thus achieved its objectives of improving energy efficiency, making the system accessible to the users and showcasing the possibility of a smart home solution that can be implemented at low costs and is controlled locally within the residence.

Although noted accomplishments have been made, there are a number of drawbacks. The prototype was built on breadboards, that are not designed for long term deployment and have wiring instability. The dashboard was running but there was some lag when using it, it was also somewhat CPU intensive, and the security was merely login / logout. No security measures were implemented to secure the data - they were just plain csv files with no security whatsoever and no good solution was provided for secure data updates, nor was there a solution for more advanced fault tolerance. Cybersecurity-wise, weaknesses in authentication, data transport encryption, and input validation were discovered, meaning that an unauthorized user can gain access into the system, tamper the data or deny access to the system.

These areas will require to be addressed in the future as the design will be integrated into a PCB for stability, scalability and the software stack will be hardened with encrypted communication and integrity protected logging. Further advanced automation concepts like predictive control based on machine learning may further optimize energy savings and yet guarantee user comfort. In practical deployment, these features of role-based access control, secure updates and anomaly detection of sensor spoofing will be crucial. All these limitations can be addressed to create a robust, secure, scalable and usable smart home system that can also be built to be resilient and sustainable.

**Acknowledgments.** I would like to express my sincere gratitude to my supervisors for their invaluable guidance, support, and feedback throughout the course of this project.